# Damping of Acoustic Waves in Straight Ducts and Turbulent Flow Conditions

H. Tiikoja, F. Auriemma, J. Lavrentjev
Tallinn University of Technology, Tallinn, Estonia

## Abstract

In this paper the propagation of acoustic plane waves in turbulent, fully developed flow is studied by means of an experimental investigation carried out in a straight, smooth-walled duct.

The presence of a coherent perturbation, such as an acoustic wave in a turbulent confined flow, generates the oscillation of the wall shear stress. In this circumstance a shear wave is excited and superimposed on the sound wave. The turbulent shear stress is modulated by the shear wave and the wall shear stress is strongly affected by the turbulence. From the experimental point of view, it results in a measured damping strictly connected to the ratio between the thickness of the acoustic sublayer, which is frequency dependent, and the thickness of the viscous sublayer of the turbulent mean flow, the last one being dependent on the Mach number. By reducing the turbulence, the viscous sublayer thickness increases and the wave propagation is mainly dominated by convective effects.

In the present work, the damping and wall impedance have been extracted from the measured complex wavenumber, which represents the most important parameter used to characterize the wave propagation. An experimental approach, referred to as iterative plane wave decomposition, has been used in order to obtain the results. The investigations have been carried out at low Mach number turbulent flows, low Helmholtz numbers and low shear wavenumbers. The aim is to overcome a certain lack of experimental results found by the authors of the most recent models for the plane wave propagation in turbulent flows, such as Knutsson et al. (The effect of turbulence damping on acoustic wave propagation in tubes, *Journal of Sound and Vibration*, Vol. 329, No. 22, 2010 4719 – 4739) and Weng et al. (The attenuation of sound by turbulence in internal flows, The *Journal of the Acoustical Society of America* 133(6) (2013) 3764–3776).

## Introduction

The propagation of plane acoustic waves in low Mach number turbulent flows is of interest for a number of commercial applications, such as: monitoring of gas pipelines by means of acoustic propagation-characteristics [1], control of the flow induced pulsation in pipe systems [2], design of thermo-acoustic devices [3], study of combustion instabilities [4].

The sound attenuation in turbulent pipe flow was first experimentally studied by Ingard & Singhal [5] who extracted the up- and downstream attenuation by using the standing-wave pattern within the duct. Moreover, the authors presented a *quasi-stationary* theory for the damping at very low acoustic frequencies. However, the experimental method used in [5] did not allow very accurate measurements of the damping. The model itself presented in [5] was a rude approximation of the reality. In fact, it neglected the influence of thermal convection and corresponded to an isothermal solution. In addition, the frequency dependence of the damping was obtained as a mere extension of the frequency dependence of the damping of waves in absence of mean flow, i.e. of the Kirchhoff model [6].

An important experimental study was performed by Ronneberger & Ahrens [7], who focused on the oscillation of the wall shear stress caused by an acoustic wave or a vibrating duct, on a turbulent boundary-layer flow. In [7], the wavenumbers were obtained by computing the ratio between the acoustic pressures measured along the duct. Since ten microphones were mounted in the pipe wall at different axial positions, the ratios resulted in an over-determined system of equations solved with regression analysis. The authors also evaluated the resonance frequency and the quality factor of a longitudinally vibrating glass pipe carrying turbulent flow. The same investigators also presented a quasi-static theory which accurately predicted the convective effects on damping at high frequency. This theory took into account the turbulent mean flow profile but it neglected the dissipation due to the interaction of the turbulent stress and the acoustic field. For a value of order one of the ratio $\delta_{ac}/\delta_l$ between the thickness of the acoustic boundary layer and the thickness of the viscous sublayer, Ronneberger & Ahrens measured a minimum in the damping at a critical Mach number. Since this minimum was not predicted by the quasi-laminar model, the authors concluded that it was due to the reflection of the shear waves generated by the acoustic waves at the edge of the viscous sublayer, which acted as a rigid plate. This assumption represented a *quasi-static* approach to the phenomenon, thus an attempt to include the sound-turbulence interaction.

In [9] Howe presented a first theoretical approach including the non-uniform turbulent eddy viscosity based on a two-dimensional flow model. This one was still a quasi-static approach which allowed predicting the global trend of the damping at high frequencies. However, the difference with the experimental results was relevant, especially at high values of the ratio $\delta_{ac}/\delta_l$. The main limitation of the quasi-static approaches was that the non-equilibrium effects introduced by the wave oscillations on the turbulent flows were neglected.

Peters et al. [9] measured the damping in a straight duct and reflection coefficient for an open pipe at low Mach and low Helmholtz number, therefore extending to lower frequencies the investigations proposed in [7]. The authors used the two microphone method described by Chung et al. [10] in order to obtain the wavenumbers in an open duct. By using more than four microphones located into two clusters – one upstream the duct and another one near to the open end – an over-determined set of non-linear equations was obtained. The three unknowns, represented by the reflection coefficient and the up- / down-stream wavenumbers were calculated by using a nonlinear regression procedure. Peters at al. also modified the rigid-plate model proposed in [9] by adding a phase shift to the reflected wave. This allowed accounting for the "memory" effects and the elastic properties of the turbulence. This modified version of the rigid-plate represented an attempt to account for the non-equilibrium effects above mentioned.

Howe proposed another model in [11] where the viscous sublayer thickness was related to the frequency via an empirical formula. In this way, the interaction mechanism between the acoustic waves and the turbulence was modeled and a satisfactory prediction of damping minima was provided. In his model, Howe replaced the boundary layer with an acoustic impedance in order to solve an inhomogeneous wave equation. Howe's model was the first complete and reliable analytical description of the damping of the acoustic waves in turbulent flows. However, it assumed uniform flow profile and it was restricted to thin acoustic boundary layers and low Mach numbers.

Allam & Åbom experimentally investigated the damping and the radiation of acoustic waves from an open duct by using the full plane wave decomposition [12]. This technique was similar to the one used by Peters et al. in [8]. Two main test cases were performed, at 100 Hz and 250 Hz, with mean flow speed up to 0.22 Mach. The authors concluded that the model proposed by Howe was reliable for a wide range of values of thickness of the boundary layer.

The simple convective model, presented by Dokumaci in [13], was accurate only for low values. However, in [14] Dokumaci presented an improved model of the Howe's model, based on the assumption of parallel sheared mean core flow instead of uniform mean core flow, as in [11]. This model still assumed negligible mean flow effects in the sublayer.

Knutsson & Åbom presented a numerical study where a finite element scheme, formulated in cylindrical coordinates, was used to account for the interaction between the turbulence and sound [15]. As a result, the authors presented a modified version of the Howe's model which included the effect of a convective mean flow. This effect was expected to become sensitive for high values of the thickness of the acoustic boundary layer. However, the complete model is still to be validated.

The models presented in [9], [11], [13] and [15], implicitly included the non-equilibrium effect due to the interactions of sound with turbulent flow. However, their theoretical foundations were still related to the quasi static assumption and the non-equilibrium effects were accounted for through semi empirical formulas. For this reason, Weng et al. in [16] and [17] presented a more complete approach where the sound-turbulence interaction process was numerically modeled, at low Mach number, by using the linearized Navier-Stokes equations (LNSE). Specifically, the turbulent stress on the sound wave, typically referred to as perturbation Reynold stress, was in focus. In this way the authors showed that the turbulence behaved like a viscoelastic fluid, instead of purely viscous, in the interaction process. This model, despite the intrinsic complexity, represents the most realistic description of the physical phenomenon of the damping in low Mach numbers turbulent flows. It allows accounting for viscothermal effects, the mean flow effects and the turbulent absorption effects. Weng et al. recently presented a work where also the mean flow refraction effects, as well as moderate compressibility effects, were tackled [18]. The refraction effects were ascribed to the non-uniform mean flow in the cross section.

Due to the limited number of experimental studied about this topic, only a few test cases have been used to validate the most recent and advanced models [15] - [17]. For this reason the main goal of the present work is to provide further experimental results about damping and the wall impedance of acoustic waves in turbulent low Mach number flows.

The paper will include a theoretical overview on the mechanism of the plane wave propagation. An experimental approach, based on iterative plane wave decompositions, will be presented. Different experimental results will be shown, in terms of damping coefficient and wall impedance measured at the flow speed range 0-0.12 Mach by using sound wave excitations at 40, 70 and 100 Hz, Helmholtz numbers 0.0153, 0.0267 and 0.0382, shear wavenumbers 84, 111 and 133 at the average temperature of 24 ℃.

Experimental results will be compared with the results provided by the models of Howe and Weng et al. in [11] and [18].

## Theoretical background

### *Damping in absence of mean flow*

At frequency of normal interest, any disturbance governed by linear equations can be considered as a superposition of vorticity, entropy and acoustic modal wave fields. The individual modal fields satisfy equations which are considerably simpler than those for the disturbance as a whole. Moreover, the equations of the model fields are uncoupled in the linear approximation, except at the boundaries. We can write:

$$V = V_{vor} + V_{ac} + V_{ent} \qquad (1)$$

for the acoustic fluid velocity. For a given angular frequency $\omega$, the dispersion relations of the vorticity and entropy modes, $k_{vor}^2 = i\omega\rho_0/\mu$ and $k_{ent}^2 = i\omega C_p/\kappa_{th}$ respectively ($\mu$ is the dynamic viscosity, $\rho_0$ is the air density and $\kappa_{th}$ is the air thermal conductivity, $C_p$ is the specific heat coefficient), are such that the imaginary part of $k$, which is typically associated with attenuation, is much larger than $\omega/c$. For this reason, since the vorticity and entropy modes are orthogonal to the velocity vector (thus, orthogonal to boundaries, interfaces and sources), those modes vanish rapidly with increasing distances from boundaries, interfaces and sources. Thus, the disturbance in an extended space is primarily made up of the acoustic-mode field, except nearby surfaces [19].

Measures of how far from boundary the vorticity and entropy mode fields extend are the respective values of $1/k_{vor}$ and $1/k_{ent}$, which represent the boundary-layer thicknesses $l_{vor}$ and $l_{ent}$:

$$l_{vor} = \sqrt{\frac{2\mu}{\omega\rho_0}}, \quad l_{ent} = \sqrt{\frac{2\kappa_{th}}{\omega\rho_0 C_p}}$$
$$= \frac{l_{vor}}{\sqrt{\Pr}} \quad (2)$$

$l_{vor}$ is also known as the thickness of the acoustic boundary layer. These lengths are not necessarily small (they tend to $\infty$ as $\omega \to 0$), but we assume that $l_{vor}$ and $l_{ent}$ are much smaller than the physical dimensions. However, in order to account for the damping of the acoustic waves in a waveguide, we recognize the presence of vorticity mode and entropy mode boundary layers. Finally, since we are interested in cases when vorticity and entropy mode fields are caused by sound of much longer wavelength than $l_{vor}$ or $l_{ent}$, we assume that these fields vary much more rapidly in the radial direction ($y$) of the waveguide than with the axial direction ($x$).

It is possible to show that, in order to satisfy the no-slip and isothermal conditions at the wall, the vorticity and entropy modes can be determined from the acoustic modal field [19]. In fact:

$$\tilde{u}_{vor,x} = -\tilde{u}_{ac,x} \exp^{-(1+i)y/l_{vor}} \quad (3)$$

$$\tilde{T}_{ent} = -\frac{1}{\rho_0 C_p}\tilde{p}_{ac} \exp^{-(1+i)y/l_{ent}} \quad (4)$$

where $\tilde{u}_{ac,x}$ and $\tilde{p}_{ac}$ are considered to be constant along the $y$ direction. For the plane waves outside the acoustic boundary layer, the solution of the perturbation equations is:

$$\tilde{p} = \tilde{p}_0 \exp^{i(\omega t - kx)} \quad (5)$$

Since the acoustic energy is dissipated when waves travel in the waveguide, the wavenumber $k$ cannot be any more equal to the real number $\omega/c_0$, as in case of neglected vorticity and entropy mode, but must be complex. In fact, as above mentioned, $\text{imag}(k) = -\alpha$, $\alpha$ being the attenuation coefficient i.e. the dissipation due to the viscous and thermal diffusions. The real part of $k$ is related to the phase velocity change.

Kirchhoff derived the wavenumber of plane waves propagating in a straight duct,

$$k_k = \frac{\omega}{c_0}\left[1 + \frac{1-i}{\sqrt{2}s}\left(1 + \frac{\gamma-1}{\xi}\right) - \frac{i}{s^2}\left(1 + \frac{\gamma-1}{\xi} - \frac{\lambda}{2}\frac{\gamma-1}{\xi^2}\right)\right] \quad (6)$$

$\gamma$ is the specific heat ratio, $\xi^2 = \mu C_p/\kappa_{th}$ is the Prandtl number, $s = a\sqrt{\rho_0 \omega/\mu}$ is the shear wavenumber and $a$ is the pipe radius. Kirchhoff's classical solution does not include the term in $s^2$, which is typically referred to as Ronneberger correction. However, this term is typically negligible [19]. Eq. (6) is valid for a homogeneous medium in absence of a mean flow.

### Damping in presence of mean flow

A turbulent flow containing coherent perturbations can be conveniently described by using the following "triple decomposition":

$$F(\mathbf{x},t) = \bar{F}(\mathbf{x},t) + \tilde{F}(\mathbf{x},t) + F'(\mathbf{x},t) \quad (7)$$

where $F$ can be the fluid density ($\rho$), temperature ($T$), the generic velocity component ($u_i$) or the pressure ($p$), $\bar{F}$ is the time average

$$\bar{F}(\mathbf{x},t) = \lim_{T\to\infty}\frac{1}{T}\int_t^{t+T} F(x,t')\mathrm{d}t' \quad (8)$$

$\tilde{F}$ is the fluctuation caused by the coherent perturbation and $F'$ is the turbulent fluctuation. Since we assume the mean flow to be stationary, $\bar{F}$ is independent of the initial time $t$ and can be considered as a function $\bar{F}(\mathbf{x})$. For periodic perturbations, $\tilde{F}$ can be extracted from the total quantity $F$ by 'phase averaging' and subtraction of the mean part $\bar{F}$. The phase averaging is defined as:

$$\langle F(\mathbf{x},t)\rangle = \lim_{N\to\infty}\frac{1}{N}\sum_{n=0}^{N} F\left(x, t + \frac{n}{f}\right) \quad (9)$$

with $f$ the frequency of the acoustic wave. In case of single frequency coherent perturbation, the phase average includes both mean flow and perturbation quantities. In measurements of acoustic waves propagating in a turbulent flow, it corresponds to the determination of the frequency response of a very noisy system, where the input is the external acoustic perturbation and the output is the quantity $\tilde{p}$.

Regarding the perturbation quantities, the coherent perturbations represented by acoustic waves have amplitudes which are sufficiently small to be well approximated by linear equations.

When writing the linearized momentum equation for the acoustic field, the quantity

$$\tilde{r}_{ij} = \langle u_i' u_j'\rangle - \overline{u_i' u_j'} \quad (10)$$

referred to as *Reynolds stress*, represents the main theoretical unknown to be modeled. It is related to the perturbation of the background Reynolds stress, $\overline{u_i' u_j'}$, due to the presence of the coherent acoustic perturbation.

Also the wave equation of the oscillating temperature $\tilde{T}$ can be derived, but from the heat equation. However, it is worth mentioning that, in this case, the perturbation of the turbulent heat flux due to the acoustic waves, $\tilde{q}_j$, appears instead of the term $\tilde{r}_{ij}$. However, $\tilde{r}_{ij}$ and $\tilde{q}_j$ are typically related to each other, for example through the gradient-diffusion hypothesis in many Reynolds Averaged Navier-Stokes (RANS) models. For this reason $\tilde{r}_{ij}$ is still of main interest.

### Main models

As a preface to this section the authors believe it is worth mentioning that the solely effect of convection in the propagation of acoustic waves in presence of a non-turbulent mean flow, is generally well described by the solution proposed by Dokumaci [14] obtained by asymptotic expansion:

$$k^\pm = \frac{\omega}{c_0} \frac{k_k}{1 \pm k_k M} \quad (11)$$

where $\pm$ denote the propagation in positive and negative $x$-direction, $k_k$ is the Kirchhoff solution and $M$ is the mean flow Mach number, expressed by Eq. (6).

However, in order to take into account the turbulent effects, the perturbation Reynolds stress must be modeled because the vorticity mode can be influenced by the turbulent stress. The turbulence can also transfer energy from the entropy mode by turbulent heat transfer.

In order to understand in which conditions the Reynolds stress and the heat transfer affect the vorticity and entropy mode, the dimensionless acoustic boundary layer thickness must be defined as in [7]:

$$\delta_A^+ = \frac{l_{vor} u_\tau}{\nu} \quad (12)$$

where $u_\tau$ is the friction velocity of the mean flow, obtainable form the relationship:

$$\frac{U_0}{u_\tau} = 2.44 \ln\left(\frac{u_\tau D_p}{2\nu}\right) + 2 \quad (13)$$

$U_0$ is the mean flow speed and $D_p$ is the hydraulic diameter. Since the thickness of the viscous sublayer of the turbulent mean flow boundary layer is $\delta_\nu \approx 5\nu/u_\tau$, $\delta_A^+$ is related to the ratio between the thickness of the acoustic boundary layer $l_{vor}$ and the thickness of the viscous sublayer $\delta_\nu$.

For high frequencies and low turbulent flows the acoustic boundary layer is thinner than the viscous sublayer, as shown in Figure 1a. In this case the turbulence does not affect the acoustic dissipation. On the contrary, for low frequencies and high turbulent flows the vorticity and entropy modes penetrate the turbulent layer and possibly also the log-law region. As a consequence, the propagations of these two modes are influenced by the turbulent mixing of momentum and heat respectively.

At this point, the main models about damping of the acoustic waves in turbulent flows can be easily understood:

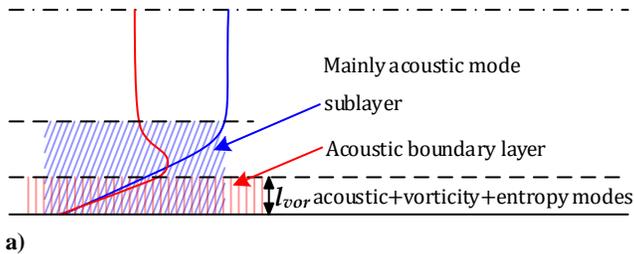

a)

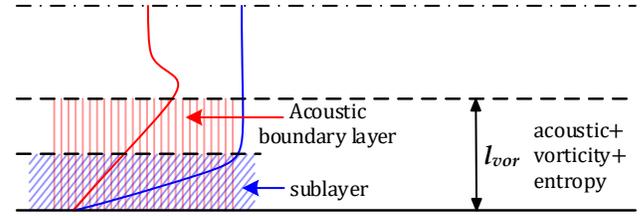

b)

Figure 1. Perturbation and mean flow profiles. a) $\delta_{ac} < \delta_l$, b) $\delta_{ac} > \delta_l$.

*Quasi-laminar models:* this is the easiest and less accurate class of models, since it neglects the presence of the fluctuation Reynolds stress, i.e. $\tilde{r}_{ij} = 0$. This means that the turbulent flow is uninfluenced by the presence of a coherent perturbation. In [7] it is shown that, at low $\delta_A^+$, these models are reliable since the shear waves decay within the viscous sublayer, therefore preventing any interaction with the turbulent flow. However, quasi-laminar models do not provide reliable results as $\delta_A^+ \geq 5$. In this case, the Reynolds stress must be somehow accounted for.

*Quasi-static models:* those models account for the sound-turbulence interaction, thus the acoustic waves modulate the Reynolds stress. However, this modulation is quasi-static in the meaning that the motion of the acoustic waves is neglected and the interaction is in equilibrium state. In these cases the Prandtl's mixing length theory can be described in order to describe the turbulence Reynolds stress by means of an eddy viscosity, i.e.:

$$\tilde{r}_{12} = -2\nu_T \frac{\partial \tilde{u}_1}{\partial x_2} \quad (14)$$

where

$$\nu_T = l_m^2 \left|\frac{\partial \bar{u}_1}{\partial x_2}\right| \quad (15)$$

is referred to as eddy viscosity of the turbulent flow and $l_m$ is the mixing length. $l_m$ is constant and independent of sound frequency in those models. Eq.s (14) and (15) allow obtaining reliable models for damping up to $\delta_A^+ \approx 10$. However, in order to predict the damping minimum at $\delta_A^+ \approx 12$, experimentally measured e.g. by Ronneberger and Ahrens in [7], several modifications of quasi-static models have been proposed by different researchers. Those modifications represent attempts to take into account the non-equilibrium effects.

*Rigid plate models:* instead of modeling the Reynolds stress, Ronneberger and Ahrens modeled the turbulence as a rigid plate which reflects the share waves generated by the sound [7]. Consequently, the authors interpreted the damping minimum at $\delta_A^+ \approx 10$ as the result of the destructive interference of the shear waves generated by the acoustic field (vorticity fields) with the shear waves reflected at the edge of the viscous sublayer by the strong variation of the eddy viscosity. For critical ratios of the acoustic boundary layer thickness and the thickness of the laminar sublayer this destructive interference is maximal, which results in the observed minimum in damping.

Ronneberger and Ahrens used an infinitely large eddy viscosity in their model, therefore overestimated the reflection of the shear wave at the edge of the viscous layer. Their model could not predict accurately the correct value of $l_{vor}/\delta_v$ where the damping minimum occurred and not even the value of this minimum.

*Implicitly included non-equilibrium models*: Peters et al. added an additional phase shift to the reflection of the "rigid plate", by introducing a relaxation time, $t_m$ which was related to the memory effect of the turbulence [9]. This is an indirect way to somehow take into account the non-equilibrium effects. In this way the minimum damping corresponds to the acoustic waves whose time period coincides with the relaxation time $t_m$. By considering this phase shift to be equal to $t_m = 100\nu/u_\tau$ the corresponding dimensionless angular frequency, also referred to as Strouhal number, $\omega^* = \omega\nu/u_\tau^2$ assumes the value $2\pi/100$, which matches the experimental values found in [7].

By introducing a frequency dependent eddy viscosity, Howe proposed a model which imposed the memory effect in the eddy viscosity $\nu_T$. For this reason this model provides a more reliable and physically realistic approach [12]. Howe proposed the following expression of the eddy viscosity:

$$\nu_T = \begin{cases} 0 & x_2 < \delta_v(\omega) \\ \kappa u_\tau |x_2 - \delta_v(\omega)| & x_2 > \delta_v(\omega) \end{cases} \quad (16)$$

where $\kappa$ is the von Karman constant ($\approx 0.41$). The frequency dependent effective thickness of the viscous sublayer $\delta_v$, was given by the following empirical formula:

$$\frac{\delta_v u_\tau}{\nu} = 6.5 \left[1 + \frac{1.7(\omega/\omega^*)^3}{1 + (\omega/\omega^*)^3}\right] \quad (17)$$

Here $\omega^*\nu/u_\tau \approx 0.01$, i.e. the relaxation time was chosen in accordance with Peters et al. in [9], as above explained ($t_m = 100\nu/u_\tau$). The model proposed by Howe is accurate for a wide range of $\delta_A^+$, although the equilibrium model provided by Weng et al., discussed in the next section, provides even more accurate physical insight. We report here the expression derived by Howe for the damping in presence of turbulent flow, since we will largely use it in comparison with our experimental results.

$$\alpha_\pm = \frac{\sqrt{2\omega}}{(1 \pm M)c_0 D_p} \mathrm{Re}\left\{\sqrt{2}\exp(-\mathrm{i}\pi/4)\left[\frac{1}{(1 \pm M)^2}\sqrt{\nu}\right.\right.$$
$$\left.\left.\times F_A\left(\sqrt{\frac{\mathrm{i}\omega\nu}{\kappa^2 u_\tau^2}}, \delta_v\sqrt{\frac{\mathrm{i}\omega}{\nu}}\right) + \frac{\beta c_0^2}{c_p}\sqrt{\chi}F_A\left(\sqrt{\frac{\mathrm{i}\omega\chi P_r^2}{\kappa^2 u_\tau^2}}, \delta_v\sqrt{\frac{\mathrm{i}\omega}{\chi}}\right)\right]\right\} \quad (18)$$

where $F_A$ is defined as

$$F_A(a,b) = \frac{\mathrm{i}(H_1^{(1)}(a)\cos b - H_0^{(1)}(a)\sin b)}{H_0^{(1)}(a)\cos b + H_1^{(1)}(a)\sin b} \quad (19)$$

and $H_0^{(1)}$ and $H_1^{(1)}$ are the Hankel functions of the zeroth and the first order respectively. In Eq. (18) $= \kappa_{th}/\rho_0 C_p$, which is for air at 20 ℃ equal to $2*10^{-5}$ [m²/s], $\beta = 1/T$ with $T$ absolute temperature.

As mentioned before, Dokumaci proposed a variant of this model which is based on the assumption of parallel sheared mean core flow [14]. Knutsson and Åbom updated the Howe's model considering the up- and down-stream propagating waves separately [15]. These models are not reported here. However, the differences with the original model by Howe are small, especially when the flow is highly turbulent.

*Non-equilibrium model:* Weng et al. presented a first complete model where the non-equilibrium effects were analytically accounted for, thus allowing a deeper understanding of the physical mechanisms involved in the sound-turbulence interaction [16] - [17]. For brevity, this model is only summarized here. In [16] a relaxation equation for the perturbation Reynolds stress was derived, where the oscillation of the deviatoric part of the Reynolds stress was expressed as a function of the perturbation share strain rate and the eddy viscosity. Unlike the traditional eddy viscosity for equilibrium flows, this parameter is a complex number depending on the frequency. As matter of fact, it can be obtained by multiplying the quasi-static eddy viscosity by a factor of $1/(1 + \mathrm{i}\omega t_\tau)$, where $t_\tau \approx 150\nu/u_\tau$. Weng et al. also showed that the proposed perturbation eddy-viscosity recovered the quasi static model at low frequencies while resembling the quasi-laminar model at high frequencies. Accordingly, the turbulence acts as a viscoelastic fluid to the perturbation and not anymore as a viscous fluid (as in non-equilibrium models). This means that the stress depends on the present strain rate but also has a memory of its past. The parameter which controls the damping is given by the ratio between the turbulent relaxation time near the wall and the sound wave period [16].

## Experiments

### Iterative wave decomposition method

In this paper, the damping of the acoustic waves propagating in turbulent flow conditions has been experimentally characterized by utilizing an iterative method based on plane wave decomposition. This method can be considered as a variant of the methods used in [7], [9] and [12].

The sound propagates in straight ducts in the form of plane waves, i.e. with a uniform wave front throughout the cross-section, as long as the frequency $f$ is below the cut-on frequency for the first non-plane mode. This frequency, in case of circular cross-sections, is given by $f_{cut-on} = 1.84c(1 - M^2)/(d\pi)$. In the frequency domain, the acoustic pressure can be written as:

$$\hat{p}(x,f) = \hat{p}_+(f)\exp(-\mathrm{i}k_+ x) + \hat{p}_-(f)\exp(\mathrm{i}k_- x) \quad (20)$$

where $\hat{p}$ is the Fourier transform of the acoustic pressure, $x$ is the axial coordinate along the duct, $\hat{p}_+$ and $\hat{p}_-$ are the wave components propagating in positive and negative $x$-direction respectively. The two-microphone technique allows calculating two unknowns among all the parameters included in Eq. (20), once the remaining parameters are known [10]. In fact, if the two microphones are placed at the generic positions 1 and 2 which are separated by the distance $s$, from Eq. (20) one obtains:

$$\begin{aligned}\hat{p}_1(f) &= \hat{p}_+(f) + \hat{p}_-(f) \\ \hat{p}_2(f) &= \hat{p}_+(f)\exp(-\mathrm{i}k_+ s) + \hat{p}_-(f)\exp(\mathrm{i}k_- s)\end{aligned} \quad (21)$$

If $s$, $k_+$ and $k_-$ are known, then $\hat{p}_+$ and $\hat{p}_-$ can be obtained. However, if $k_+$ and $k_-$ are also unknown it is necessary to add other two equations to Eq. (21) in order to solve the system, i.e. other two microphones are required for the measurements. The results obtained by using four microphones for the determination of those unknowns can be sensitively improved by increasing the number of microphone positions, which results into an over-determined system of equations. For a set of $n$ microphone locations $x_j$ ($j = 1 \ldots n$) the following system is obtained:

$$\begin{bmatrix} \exp(-ik_+x_1) & \exp(ik_-x_1) \\ \exp(-ik_+x_2) & \exp(ik_-x_2) \\ \ldots & \ldots \\ \ldots & \ldots \\ \exp(-ik_+x_n) & \exp(ik_-x_n) \end{bmatrix} \begin{bmatrix} \hat{p}_+ \\ \hat{p}_- \end{bmatrix} = \begin{bmatrix} \hat{p}_1 \\ \hat{p}_2 \\ \ldots \\ \ldots \\ \hat{p}_n \end{bmatrix} \quad (22)$$

In order to solve the system in Eq. (22) the vectorial function , whose components are defined as

$$f_j = \hat{p}_+ \exp(-ik_+x_j) + \hat{p}_- \exp(ik_-x_j) - \hat{p}_j \quad j = 1 \ldots n \quad (23)$$

must be minimized. In Eq. (23) $\hat{p}_j$ is the acoustic pressure measured by the microphone placed at the generic location $x_j$. The function $\boldsymbol{f}$ can be minimized according to different methods. Among the others we mention the method of weighted residuals, the Gauss-Newton iterative procedure, the Levenberg-Marquardt algorithm. In the present work, the Levenberg-Marquardt algorithm has been utilized and implemented by means of Matlab® package.

The pairs of unknowns ($\hat{p}_+$, $\hat{p}_-$) and ($k_+$ and $k_-$) in Eq. (23) $k_+$ have different order of magnitudes and small variations of the first pair result in sensitive variations of the second one. On the other hand, sensitive changing of ($k_+$ and $k_-$) can have minor effects on the corresponding ($\hat{p}_+$, $\hat{p}_-$). An over-determined system, such as Eq. (22), always involves the presence of an error in the solutions and the error on ($\hat{p}_+$, $\hat{p}_-$) can threat the reliability of the solution for ($k_+$ and $k_-$). Moreover, $k_+$ and $k_-$ will strongly depend on the number and location of microphones used for the over-determination and the results are affected by uncertainty.

For this reason, instead of solving Eq. (22) with four unknowns, Eq. (21) has been solved by assuming initial guess values of $k_+$ and $k_-$ provided by Eq. (11). In this way, an initial but quite reliable estimation of ($\hat{p}_+$, $\hat{p}_-$) is obtained, which is then used to solve Eq. (22) with the only unknowns $k_+$ and $k_-$. The new values obtained for $k_+$ and $k_-$ are used to solve once again Eq. (21). The process can be iterated several times. Usually after 5-6 iterations the solution converges with accurate estimate of all the four parameters.

*Description of the measurement setup*

A sketch of the experimental facility used to determine the damping of the propagating acoustic wave can be seen in Figure 2. The measurements have been carried out by using circular stainless steel ducts with inner radius of $a = 0.021$ m, wall thickness of $l = 0.0015$ m and wall roughness less than 1 μm. To minimize disturbances, the test-section has been supported with rubber dampers and connected with acoustic and flow sources through ALFAGOMMA T-634 rubber hose. The flow, up to $U = 50$ m/s measured at the duct centerline, has been generated by a centrifugal blower (Kongskilde 300TRV) driven by a 22 kW electric motor. The flow velocity, pressures and temperatures have been determined by a portable anemometer (Delta Ohm HD 2114P.0 and HD 2164.0) using a Pitot tube mounted in the duct. The sine wave acoustic excitation has been provided by Hertz EV165S loudspeakers connected with 0.05 m perforated side branches to reduce flow instabilities. The loudspeakers have been driven by a software-based signal generator through NI9269 analog output module and power amplifier (Velleman VPA2100MN). The signal acquisition has been performed by a dynamic signal analyzer (NI PCI-4474 and NI 9234), controlled by PC based virtual instrument (LabVIEW®).

The acoustic pressures have been measured by six flush mounted ¼ inch prepolarized G.R.A.S. pressure microphones 40BD together with 26CB preamplifiers. To measure the damping of the sound wave, the microphones have been positioned into two clusters separated by the distance of $s_1 = 2.49$ m. To reduce sensitivity to the errors, microphone separation $s_2 = 0.8$ m in both clusters has been chosen according to [11].

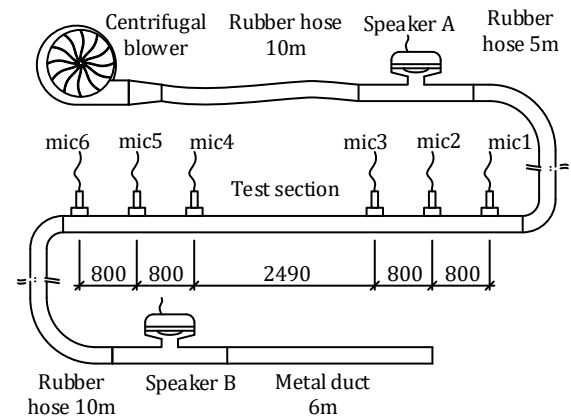

Figure 2. A sketch of the experimental set-up.

## Experimental results

By using the iterative wave decomposition when the loudspeaker A and loudspeaker B alternately play, the wavenumbers $k_+$ and $k_-$ and, as a result, the damping coefficients $\alpha_+$ and $\alpha_-$ have been determined as a function of the Mach number at constant frequency cases. The Table I shows the flow data at every frequency case. It can be noticed that a quite sensitive temperature shift affects the measurements due to the air compression taking place within the centrifugal blower.

Figures 3, 4 and 5 show the damping coefficients obtained by using constant Helmholtz numbers, corresponding to 40 Hz, 70 Hz and 100 Hz respectively, as a function of the dimensionless acoustic boundary layer thickness $\delta_A^+$ defined by Eq. (12). In fact, as above mentioned, $\delta_A^+$ is the key factor for this type of investigation, since it is related to the ratio between the thickness of the acoustic boundary layer $l_{vor}$ and the thickness of the viscous sublayer $\delta_\nu$. In the figures the damping coefficients are normalized by the values $\alpha_0$ calculated by using Eq. (6), which is a function of the temperature.

Table 1. Physical properties of the experimental flow cases studied.

| Flow case | $U_{cl}$ [m/s] | $T$ [°C] | $c$ [m/s] | $M$ |
|---|---|---|---|---|
| 1 | 0 | 21 | 343 | 0 |
| 2 | 3.5 | 21.2 | 344 | 0.0083 |
| 3 | 7.1 | 21.2 | 344 | 0.0169 |
| 4 | 11.1 | 21.7 | 344.3 | 0.0264 |
| 5 | 15.3 | 22.8 | 344.9 | 0.0364 |
| 6 | 19.6 | 23.8 | 345.5 | 0.0465 |
| 7 | 24.1 | 25.9 | 346.7 | 0.057 |
| 8 | 28.7 | 27.4 | 347.6 | 0.0677 |
| 9 | 33.2 | 30.5 | 349.4 | 0.0779 |
| 10 | 38 | 35.3 | 352 | 0.0885 |
| 11 | 42.8 | 38.8 | 354 | 0.0991 |
| 12 | 47.6 | 41.5 | 355.5 | 0.11 |
| 13 | 52.5 | 47.5 | 358.8 | 0.12 |

The experimental results presented here confirm that, in presence of a mean flow, different damping coefficients are exhibited by the acoustic waves which travel in upstream and in downstream direction. This is related to the convective effect of the flow. For small values of $\delta_A^+$, the damping of the acoustic waves is not influenced by the turbulent stress. However, at the low frequencies examined here, the acoustic boundary layer is thicker than the viscous sublayer and a significant damping is shown. For a value of this ratio of order two (which corresponds to $\delta_A^+ \approx 10$), at critical Mach number a minimum in the damping is observed due to the destructive interference of the shear waves generated by the acoustic field and the shear waves reflected at the edge of the viscous sublayer by the strong variation of the eddy viscosity.

The Figures 3-5 show that the model presented by Howe is in general agreement with the experimental results. In particular, the damping minima are captured well in terms of value and position. The collapse between the modeled and measured damping is good in the downstream direction and mainly qualitative in the upstream direction. In fact, the predicted damping coefficients in the upstream case are always overestimated. In Figure 5, the measured damping is compared with the predictions by Weng et al. [16] - [17] and with the experimental values obtained by Allam et al. [12] at 20 °C. The experimental damping coefficients in downstream direction are in good agreement among each other and with the predicted values. However, a certain disagreement is noticeable in the upstream damping coefficients also among the different experimental approaches. The results from the present investigation are affected by $\Delta T = 26.5$ °C temperature shift between the flow cases at 0 and at 0.12 Mach number. Although the experimental results and the predictions obtained by using the Howe's model are normalized by using values $\alpha_0$ calculated at the measured temperatures, a $\Delta T$ of this entity can results in a difference of 6 % on the normalized damping coefficients. Once accounted this aspect, it can be noticed that the measured damping is still closer than the damping obtained by Allam et al. to the numerical predictions of Weng et al. Moreover, the theoretical values always appear in between the two different experimental results. Further investigations are currently in progress to determine the presence of possible measurement errors in this – and the other – test cases, which can be related to the presence of pressure minima at one or more microphone locations.

In Figures 6 and 7, the real and imaginary part of the averaged wall impedance,

$$Z = \lim_{M \to 0} \frac{\bar{k} - (\omega/c_0)/(1-M^2)}{-i\bar{\alpha}} \quad (24)$$

are plotted. In Eq. (24) $\bar{k} = (k_+ + k_-)/2$ and $\bar{\alpha} = (\alpha_+ + \alpha_-)/2$. $Z$ represents the weighted sum of the shear stress impedance $Z_\tau$ and the heat conduction impedance $Z_q$. $Z_\tau$ and $Z_q$ are respectively the ratio of the complex amplitudes of the shear stress and the velocity in the shear wave and the ratio of the conducted heat and the temperature in the heat conduction wave, the ratios being evaluated at the wall. The data have not been extrapolated to zero Mach number, but for low Mach numbers the difference between $k_0$ and $\lim_{M \to 0} k_\pm$ is of the order of $M^2$. The real part of $Z$ is equal to the normalized average damping coefficient $(\bar{\alpha}/\alpha_0)$. In the range of $\delta_A^+$ from 10 to 15 the damping slightly decreases, which shows that in this situation the damping is lower than in absence of flow.

A similar trend is found for the imaginary part of $Z$, which is related to the change of the phase velocity of the wave, see Figure 7. Here, for $\delta_A^+$ lower than 23 the values extracted from the three cases always decrease. For higher values of $\delta_A^+$ those values start to increase.

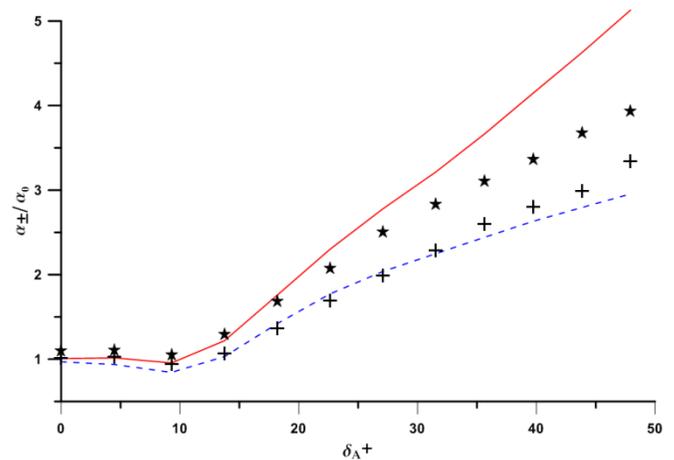

Figure 3. Damping coefficient $\alpha_\pm/\alpha_0$ as a function of $\delta_A^+$ at $ka = 0.0153$ (40 Hz). + measured downstream, ★ measured upstream, --- predicted by Howe's model downstream, —— predicted by Howe's model upstream.

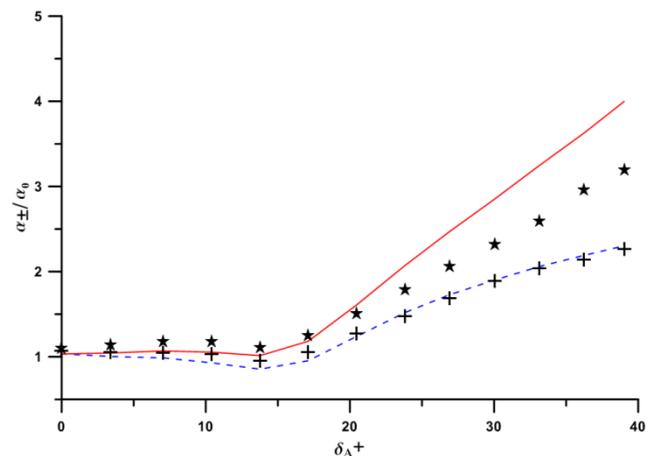

Figure 4. Damping coefficient $\alpha_\pm/\alpha_0$ as a function of $\delta_A^+$ at $ka = 0.0267$ (70 Hz). + measured downstream, ★ measured upstream, --- predicted by Howe's model downstream, —— predicted by Howe's model upstream.

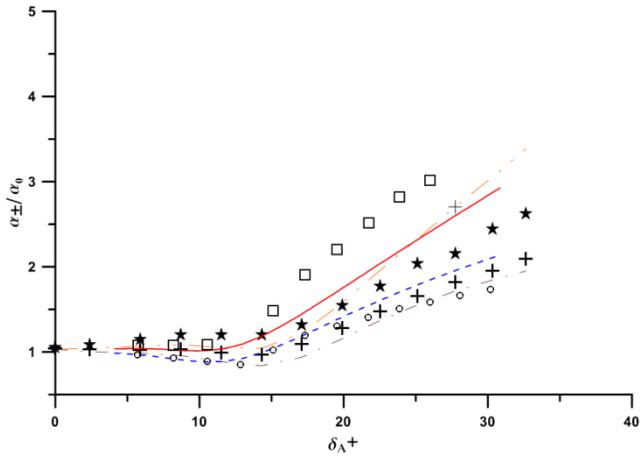

Figure 5. Damping coefficient $\alpha_{\pm}/\alpha_0$ as a function of $\delta_A^+$ at $ka = 0.0267$ (100 Hz). + measured downstream, ★ measured upstream, --- predicted by Weng's model downstream, —— predicted by Weng's model upstream. Results from [12]: □ upstream and ○ downstream. –·–·– Predicted by Howe downstream and –··–··– predicted by Howe upstream.

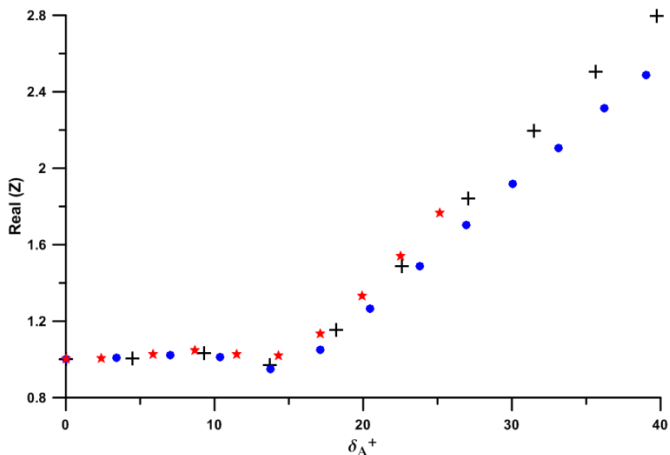

Figure 6. Real part of the averaged wall impedance as a function of $\delta_A^+$ at: + $ka = 0.0153$ (40 Hz), ● $ka = 0.0267$ (70 Hz) and ★ $ka = 0.0382$ (100 Hz).

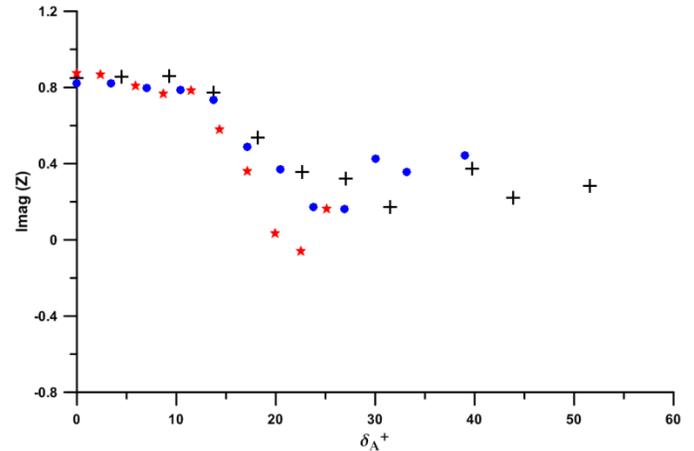

Figure 7. Imaginary part of the averaged wall impedance as a function of $\delta_A^+$ at: + $ka = 0.0153$ (40 Hz), ● $ka = 0.0267$ (70 Hz) and ★ $ka = 0.0382$ (100 Hz).

## Conclusions

In this work the damping of the acoustic plane waves in presence of turbulent flow has been experimentally characterized at low Helmholtz numbers and Mach numbers ranging from $0$ to $0.12$. The frequencies investigated are $40$, $70$, 100 Hz.

The experimental results have been carried out by utilizing an iterative method based on plane wave decomposition. In this way, the pairs of unknowns, $(p^+, p^-)$ and $(k^+, k^-)$, have been extracted separately and iteratively, once a first guess of $(k^+, k^-)$ is provided.

Different theoretical approaches have been summarized and the experimental results have been compared to the values predicted by Howe and Weng et al. A good agreement between experimental and theoretical values has been found in both cases. However, the model by Howe always overestimates the damping in the upstream direction. The model presented by Weng et al. has shown values of damping in upstream direction which are intermediate between the experimental results obtained in this work and the ones obtained by Allam et al. in [12]. This discrepancy is currently under investigation.

The real and the imaginary part of the wall impedance have been extracted by using the average wavenumbers. The first part is related to the averaged damping coefficient which, at certain values of the normalized acoustic boundary layer, can be lower than the in absence of mean flow. The imaginary part, related to the phase shift velocity of the wave, noticeably decreases up to values of the normalized acoustic layer of 20-30, but a certain recover of is shown for higher values.

## References

1. Liu C.W., Li Y.X., Fu J.T., Liu G.X., "Experimental study on acoustic propagation-characteristics-based leak location method for natural gas pipelines", Process Safety and Environmental Protection 9 6 (2015) 43–60, doi: http://dx.doi.org/10.1016/j.psep.2015.04.005.
2. D. Tonon, A. Hirschberg, J. Golliard, S. Ziada, "Aeroacoustics of pipe systems with closed branches", International Journal of

Aeroacoustics, Vol. 10(2011), No. 2-3, p. 201-276, doi: http://dx.doi.org/10.1260/1475.472X.10.2-3.201.

3. Babaei H., Siddiqui K., "Design and optimization of thermoacoustic devices", Energy Conversion and Management 49 (2008) 3585–3598, doi:10.1016/j.enconman.2008.07.002.
4. O'Connor J., 1, Acharya V., Lieuwen T., "Transverse combustion instabilities: Acoustic, fluid mechanic, and flame processes", Progress in Energy and Combustion Science 49 (2015) 1e39 doi: http://dx.doi.org/10.1016/j.pecs.2015.01.001.
5. Ingard U., Singhal V.K., "Sound attenuation in turbulent pipe flow", Journal of Acoustic Society of America, V ol. 55, No.3, March 1974.
6. Kirchhoff G., "Uber den Einfluss der Wärmteleitung in einem Gase auf die Schallbewegung", Pogg. Ann. Phys. 134 (6), 177-193, 1868.
7. Ronneberger D., Ahrens C., "Wall shear stress caused by small amplitude perturbations of turbulent boundary-layer flow : an experimental investigation", Journal of Fluid Mechanics 83, 433-464, 1977.
8. Howe M. S., "On the absorption of sound by turbulence and other hydrodynamic flows", IMA Journal of Applied Maths (1984), vol. 32, pp 187-209.
9. Peters M. C. A. M., Hirshberg A., Reijnen A.J., Wijnands A.P., "Damping and reflection coefficient measurements for an open pipe at low Mach number and low Helmholtz number", Journal of Fluid Mechanics (1993), vol. 256, pp.499-534.
10. Chung J.Y, Blaser D.A, "Transfer function method of measuring in-duct acoustic properties. I. Theory", Journal of. Acoustic Society of America 68, 907 (1980), doi:http://dx.doi.org/10.1121/1.384778
11. Howe M. S., "The damping of sound by wall turbulent shear layers", Journal of Sound and Vibration , 98(3): 1723-1730, 1995.
12. Allam S., Åbom M., "Investigation of damping and radiation using full plane wave decomposition in ducts", Journal of Sound and Vibration 292 (2006), 519-534, doi: 10.1016/j.jsv.2005.08.016
13. Dokumaci E., "A note on transmission of sound in a wide pipe with mean flow and viscothermal attenuation", Journal of Sound and Vibration 208 (4) (1997), 653-655.
14. Dokumaci E., "On attenuation of plane sound waves in turbulent mean flow", Journal of Sound and Vibration, 320 (2009)1131–1136.
15. Knutsson, M., Åbom, M., "The effect of turbulence damping on acoustic wave propagation in tubes", Journal of Sound and Vibration, Vol. 329, No. 22, 2010, pp. 4719 – 4739, doi:10.1016/j.jsv.2010.05.018.
16. Weng C., Boij S., Hanifi A., "The attenuation of sound by turbulence in internal flows", The Journal of the Acoustical Society of America 133(6)(2013) 3764–3776, doi:http://dx.doi.org/10.1121/1.4802894.
17. Weng C., Boij S., Hanifi A., "On the calculation of the complex wavenumber of plane waves in rigid-walled low-Mach-number turbulent pipe flows", Journal of Sound and Vibration (2015), doi:http://dx.doi.org/10.1016/j.jsv.2015.06.013i.
18. Weng C., Boij S., Hanifi A., "Sound-turbulence interaction in low Mach number duct flow", 19th AIAA/CEAS Aeroacoustics Conference. 2013.
19. Pierce A. D., "Acoustics : an introduction to its physical principles and applications", The Acoustical Society of America, 1991. ISBN 0070499616 0070499624.
20. Bodén H., Åbom M., "Influence of errors on the two-microphone method for measuring acoustic properties in ducts", Journal of Acoustic Society of America, (1986) vol. 79, pp.541-549.

## Contact Information

Heiki Tiikoja, heiki.tiikoja@ttu.ee
Fabio Auriemma, fabio.auriemma@ttu.ee

## Acknowledgments

The research was supported by Innovative Manufacturing Engineering Systems Competence Centre IMECC and Enterprise Estonia (EAS) and co-financed by European Union Regional Development Fund project EU48685.